# Feature based Sequential Classifier with Attention Mechanism

Sudhir Sornapudi, R. Joe Stanley, William V. Stoecker, Rodney Long, Zhiyun Xue, Rosemary Zuna, Shelliane R. Frazier, Sameer Antani

*Abstract*—**Cervical cancer is one of the deadliest cancers affecting women globally. Cervical intraepithelial neoplasia (CIN) assessment using histopathological examination of cervical biopsy slides is subject to interobserver variability. Automated processing of digitized histopathology slides has the potential for more accurate classification for CIN grades from normal to increasing grades of pre-malignancy: CIN1, CIN2 and CIN3. Cervix disease is generally understood to progress from the bottom (basement membrane) to the top of the epithelium. To model this relationship of disease severity to spatial distribution of abnormalities, we propose a network pipeline, DeepCIN, to analyze high-resolution epithelium images (manually extracted from whole-slide images) hierarchically by focusing on localized vertical regions and fusing this local information for determining Normal/CIN classification. The pipeline contains two classifier networks: 1) a cross-sectional, vertical segment-level sequence generator (two-stage encoder model) is trained using weak supervision to generate feature sequences from the vertical segments to preserve the bottom-to-top feature relationships in the epithelium image data; 2) an attention-based fusion network image-level classifier predicting the final CIN grade by merging vertical segment sequences. The model produces the CIN classification results and also determines the vertical segment contributions to CIN grade prediction. Experiments show that DeepCIN achieves pathologist-level CIN classification accuracy.**

This research was supported in part by the Intramural Research Program of the National Institutes of Health (NIH), National Library of Medicine (NLM), and Lister Hill National Center for Biomedical Communications (LHNCBC).

S. Sornapudi is with the Department of Electrical and Computer Engineering of Missouri University of Science and Technology in Rolla, MO 65409-0040 USA (e-mail: ssbw5@mst.edu)

R. J. Stanley is with the Department of Electrical and Computer Engineering of Missouri University of Science and Technology in Rolla, MO 65409-0040 USA (e-mail: stanleyj@mst.edu).

William V. Stoecker is with Stoecker & Associates, Rolla MO, 65401, USA (e-mail: wvs@mst.edu).

R. Long is with the Lister Hill National Center for Biomedical Communications for National Library of Medicine, National Institutes of Health, Bethesda, MD 20894, USA (e-mail: long@nlm.nih.gov)

Z. Xue is with the Lister Hill National Center for Biomedical Communications for National Library of Medicine, National Institutes of Health, Bethesda, MD 20894, USA (e-mail: zhiyun.xue@nih.gov)

R. Zuna is now retired. During this effort she was with the Department of Pathology for the University of Oklahoma Health Sciences Center in Oklahoma City, OK 73117, USA (email: rosemary-zuna@ouhsc.edu)

Shellaine R. Frazier is with the Surgical Pathology Department for the University of Missouri Hospitals and Clinics in Columbia, MO 65202, USA (email: FrazierSR@health.missouri.edu)

S. Antani is with the Lister Hill National Center for Biomedical Communications for National Library of Medicine, National Institutes of Health, Bethesda, MD 20894, USA (email: sameer.antani@nih.gov)

## I. Introduction

Cervical cancer prevention remains a big global challenge. It is estimated that in 2020 in the US 13,800 women will be diagnosed with invasive cervical cancer, and among them 4,290 will die [1]. This cancer ranks second in fatalities among 20-39 year old women [1]. Screening has helped decrease the incidence rate of cervical cancer by more than half since the mid-1970s through early detection of precancerous cells [2], yet 300,000 women die every year worldwide [3]. As a public health priority in 2018 the WHO director general made a global call for elimination of cervical cancer [4].

If clinically indicated, the cervix is further examined by taking a sample of cervical tissue (biopsy). The tissue sample is transferred to a glass slide and observed under magnification (histopathology). Cervical dysplasia or cervical intraepithelial neoplasia (CIN) is the growth of abnormal cervical cells in the epithelium that can potentially lead to cervical cancer. CIN is usually graded on a 1-3 scale. CIN 1 (Grade I) is mild epithelial dysplasia, confined to the inner one third of the epithelium. CIN 2 (Grade II) is moderate dysplasia, usually spread within the inner two-thirds of the epithelium. CIN 3 (Grade 3) is carcinoma in-situ (severe dysplasia) involving the full thickness of the epithelium [5]. A diagnosis of Normal indicates the absence of CIN. Fig. 1 depicts the localized regions with all four classes.

Our previous work on computational approaches for digital pathology image analysis has relied mostly on extraction of handcrafted features based on the domain expert's knowledge. Guo *et al.* [6] manually extracted traditional nuclei features for CIN grade classification. The images were split into ten equal vertical segments for extraction of local features, and classified using voting fusion with support vector machine (SVM) and linear discriminant analysis (LDA). Huang *et al.* [7] used the LASSO algorithm for feature extraction with SVM ensemble learning for classification of cervical biopsy images. Automated CIN grade diagnosis was also performed through analyzing Gabor texture features with K-means clustering [8] and slide level classification with texture features [9]. Accuracy fell short of that needed for clinical or laboratory use. In the past decade, success of deep learning approaches for image segmentation and classification in the health domain has attracted more research [10]. Toward that, AlMubarak *et al.* [11] developed a fusion-based hybrid deep learning approach that combined manually extracted features and convolutional neural network (CNN) features to detect the CIN grade from histology images. Li *et al.* [12] proposed a transfer learning framework with the Inception-v3 network for classifying cervical cancer images. An excellent review of computer vision approaches for cervical histopathology image analysis was presented in Li *et al.* [13].

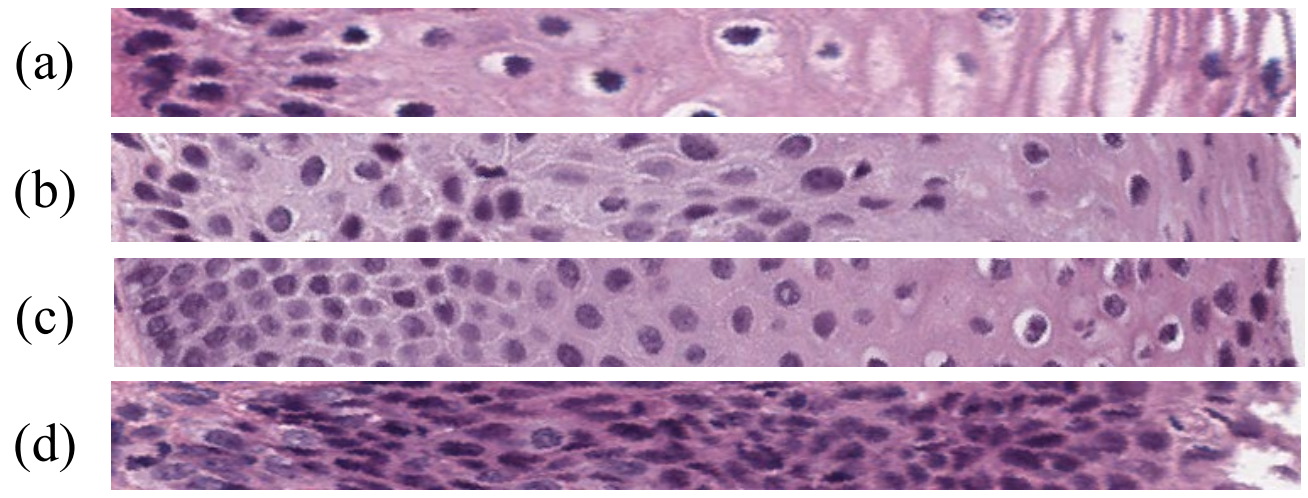

Fig. 1. Sections of epithelium region with increasing CIN severity (from (b)-(d)) showing delayed maturation with increase in immature atypical cells from bottom-to-top. The sections can be categorized as (a) Normal, (b) CIN1, (c) CIN2, and (d) CIN3. In these images left-to-right corresponds to bottom-to-top of the epithelium.

A critical problem with manual CIN grading by pathologists is the variability among general pathologists in CIN determination. Stoler *et al.* [14] found agreement for the general community pathologist with the expert pathologist panel assignment to range from 38% to 68%: 38.2%, 38%, and 68% for CIN grades 1, 2 and 3, respectively. The overall Cohen's kappa value (κ) was 0.46 for four grades, these three CIN grades and cervical carcinoma. Cai *et al.* [15] found close agreement among expert pathologists. For four expert pathologists, with 8-30 years of grading CIN slides, a weighted κ range of 0.799 to 0.887 was found. If automated CIN grading results can be made as close to expert readings as the variability among expert pathologist readings, automated CIN grading may become feasible.

Our proposed DeepCIN pipeline draws inspiration from the way pathologists examine epithelial regions under the microscope. They do not scan the entire slide at once, instead they analyze local regions across the epithelium to understand the bottom-to-top growth of atypical cells and to compare the relative sizes of the cell nuclei in local neighborhoods. They use this local information to decide the CIN grade globally for the whole epithelium region. We developed a pathologist-inspired automated pipeline analogous to human study of histopathology slides, where we first localize the epithelial regions, then we analyze the features across these regions in both directions; finally, we fuse the feature information to predict the CIN class label and estimated the contribution of these local regions towards the global class result.

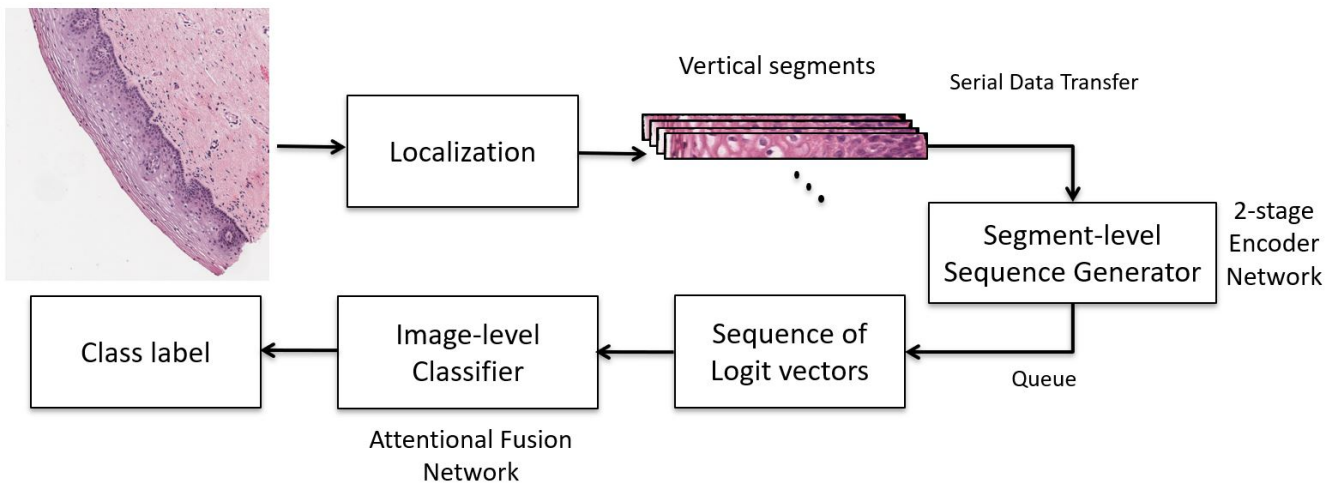

Fig. 2. Overview of DeepCIN model.

In this paper, we present DeepCIN, to automatically categorize high-resolution cervical histology images into Normal or one of the three CIN grades. Images used in this work are manually segmented epithelium regions extracted from digitized whole slide images (WSIs) at 10X magnification. The classification is carried out through hierarchical analysis of local epithelial regions by focusing on individual vertical segments and then combining the localized feature information in spatial context by introducing recurrent neural networks (RNNs).

The use of RNNs [16], [17] has been found to be successful in solving time-series and sequential prediction problems. Their use has led to better understanding of contextual features from images when combined with CNN-based models. Typically, CNNs act as a feature extractor, and RNNs learn the contextual information. Shi *et al.* [18] proposed a convolutional recurrent neural network for scene text (sequence-to-sequence) recognition. Attention mechanisms [19] were incorporated later to improve performance [20], [21]. Attention-based networks have been used in speech, natural language processing, statistical learning and computer vision [22].

A key aspect of our model is that it focuses on differentially informative vertical segment regions. This is crucial for deciding the level of CIN, because variation of CIN grade in local region could impact the overall CIN assessment of the epithelium [23]. The major contributions of this paper are:

1) Hierarchical image analysis from localized regions to the whole epithelium image.
2) Capturing the varying nuclei density across the epithelium region by vertically splitting the region into standard width segments with reference to the medial axis.
3) Weakly supervised training scheme for vertical segments.
4) Image-to-sequence two-stage encoder model for extracting localized segment level information.
5) Attention-based fusion (many-to-one model) for whole epithelium image CIN classification.
6) Identifying local segment contributions towards the whole image CIN classification.

## II. Methodology

DeepCIN incorporates a two-fold learning process (Fig. 2). First, generated vertical segments from the epithelial image are fed to a two-stage encoder model for weak supervision training

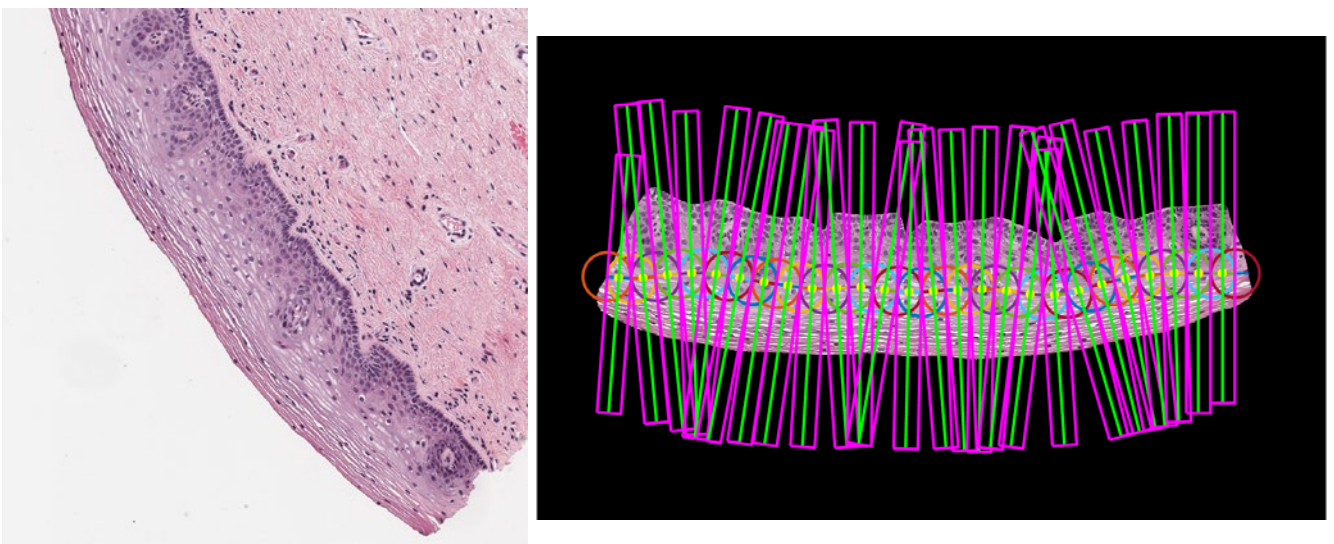

Fig. 3.  Localized vertical segment generation from an epithelial image.

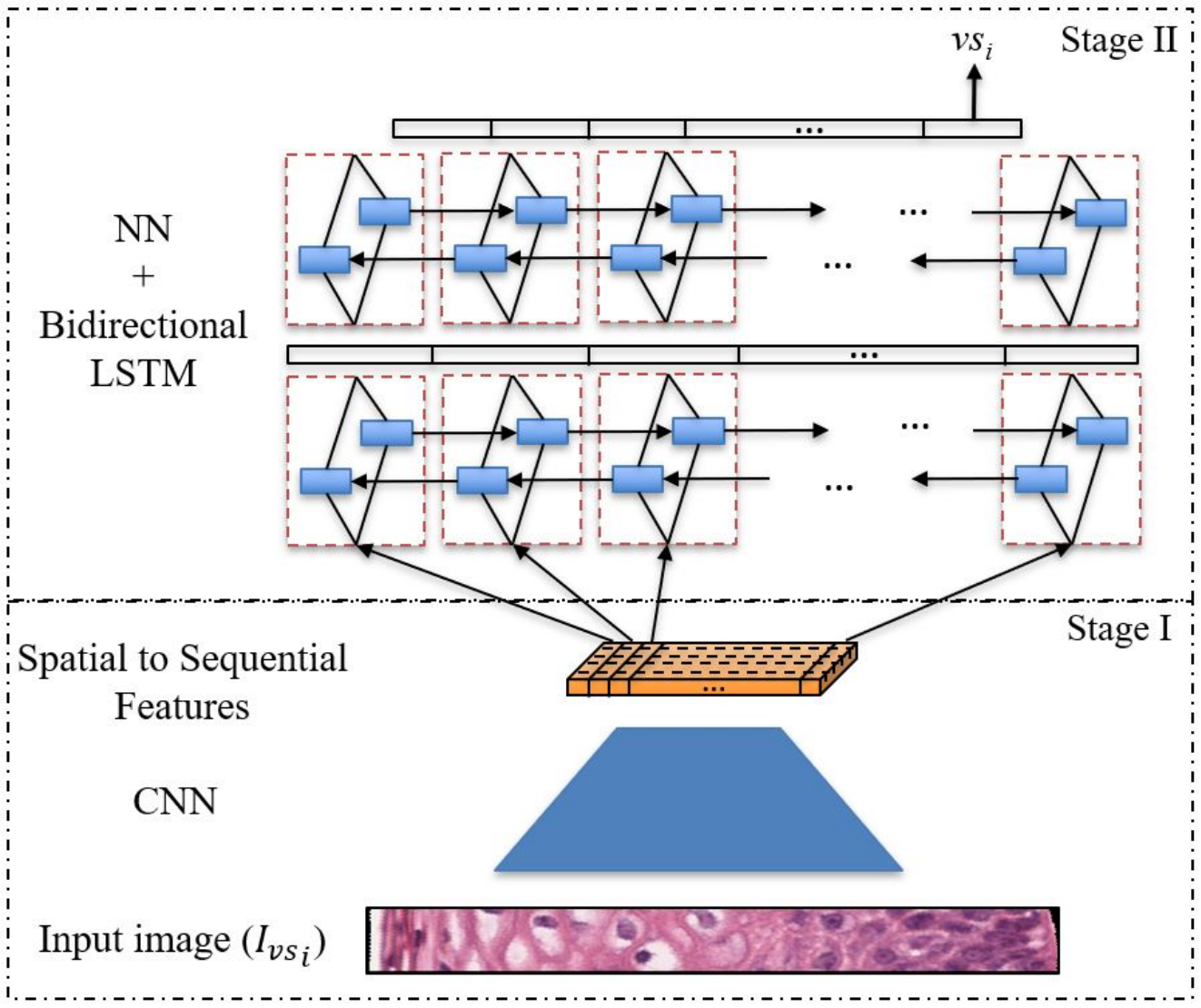

Fig. 4.  Segment-level sequence generator network with two-stage encoder structures.

TABLE I
SEGMENT-LEVEL SEQUENCE GENERATOR MODEL ARCHITECTURE

| | Layers | Configurations | Size |
|---|---|---|---|
| Stage I | Input | - | $3 \times 64 \times 704$ |
| | Transition Layer 0 | $k: 7 \times 7, s: 2, p: 3$ | $64 \times 32 \times 352$ |
| | | $mp: 3 \times 3, s: 2, p: 1$ | $64 \times 16 \times 176$ |
| | Dense Block 1 | $\begin{bmatrix} k: 1 \times 1, s: 1, p: 1 \\ k: 3 \times 3, s: 1, p: 1 \end{bmatrix} \times 6$ | $256 \times 16 \times 176$ |
| | Transition Layer 1 | $\begin{bmatrix} k: 1 \times 1, s: 1 \\ ap: 2 \times 2, s: 2 \end{bmatrix}$ | $128 \times 8 \times 88$ |
| | Dense Block 2 | $\begin{bmatrix} k: 1 \times 1, s: 1, p: 1 \\ k: 3 \times 3, s: 1, p: 1 \end{bmatrix} \times 12$ | $512 \times 8 \times 88$ |
| | Transition Layer 2 | $\begin{bmatrix} k: 1 \times 1, s: 1 \\ ap: 2 \times 2, s: 2 \end{bmatrix}$ | $256 \times 4 \times 44$ |
| | Dense Block 3 | $\begin{bmatrix} k: 1 \times 1, s: 1, p: 1 \\ k: 3 \times 3, s: 1, p: 1 \end{bmatrix} \times 24$ | $1024 \times 4 \times 44$ |
| | Pooling | $mp: 3 \times 3, s: 4 \times 1$ | $1024 \times 1 \times 44$ |
| Stage II | BLSTM + NN | $nh: 256$ | $512 \times 44$ |
| | | $nh: 256$ | $256 \times 44$ |
| | BLSTM + NN | $nh: 256$ | $512 \times 44$ |
| | | $nh: 4$ | $4 \times 44$ |
| | Output | - | $4 \times 1$ |

$k$, $s$, $p$, $mp$, $ap$, and $nh$ are kernel, stride size, padding size, max pooling, average pooling and number of hidden layers, respectively. 'BLSTM' and 'NN' stands for bi-directional LSTM and single layer neural network, respectively.

to constrain the segment class to the image class. Second, an attention-based fusion network is trained to learn the contextual feature information from the sequence of segments and classify the epithelial image into one of the four classes. The remainder of this section of the paper is organized as follows: Section II.A discusses cross-sectional vertical segment generation within an epithelium image; Section II.B and Section II.C present the two parts of the model: a segment-level sequence generator and an image-level classifier; Section II.D describes the model training approach.

### *A. Localization*

Initially, we process the manually segmented epithelium regions to find the medial axis and reorient the epithelium to be aligned horizontally, as performed by Guo *et al.* [6]. Guo's methods are modified to generate standard-width vertical segments with reference to the medial axis. This helps in better understanding the pattern of atypical cells under uniform epithelium sections and generating more image data for training our deep learning model. We approximate the medial axis curve as a piece-wise linear curve by iteratively drawing a series of circles (left to right) of radii equal to the desired segment width. The center of each successive circle is the right-most intersection point of the previously drawn circle and the medial axis curve. All the consecutive intersection points along the medial axis curve are joined to form a polygonal chain. At the midpoint of each line segment, we compute the slope corresponding to an intersecting perpendicular line. At the end points of the line segment, we draw vertical lines parallel to this midpoint perpendicular. This creates rectangular vertical regions of interest as shown in Fig. 3. Using these individual vertical regions, we compute a bounding box, which we apply to the original image to crop a refined vertical segment. The heights and counts of vertical segments created in this manner vary with the shapes and sizes of the epithelial images. The height and width of the segments are empirically chosen to be 704 pixels and 64 pixels, respectively (for details refer to Section III.A). The RGB image segments are further processed by channel-wise normalizing the pixel intensities with zero mean and standard deviation of value one, and rotating counter-clockwise by 90 degrees. This facilitates the classification of localized epithelial regions.

Formally, we assume that an epithelial image $I_{epth}$ has $N$ vertical segments $I_{vs_i}$ such that:

$$I_{epth} = \{I_{vs_1}, I_{vs_2}, \ldots, I_{vs_N}\}. \quad (1)$$

### *B. Segment-level Sequence Generation*

The segment-level sequence generator network is built as an two-stage encoder classifier model. The main objective of this network is to generate logit vectors to serve as localized sequence information for further image-level analysis. Because ground-truth labels for our vertical segments are not available, the network is trained against the image-level CIN grade. Since we expect variability in the true CIN grades across the vertical segments, use of the single image-level grade for all segments within an image introduces noisy labelling for the segments, and this may be expected to affect our training. Hence, we consider this a weakly supervised learning process.

We tackle this classification problem as a sequence recognition problem. As shown in Fig. 4, the stage I encoder is constructed with a CNN that can extract the convolutional feature maps. These spatial features are then reduced to have height of 1 with maximum pooling operation. It is further transformed into a feature sequence by splitting along its width and concatenation of vectors formed by joining across the channels, similar to Shi *et al.* [18]. The RNN acts as a stage II encoder that decodes the sequential information to predict the class value (many-to-one model). It is important to understand that the vertical segments carry valuable localized feature information including varying nuclei density, which is crucial in the decision process. Therefore, it is well represented as a feature sequence and a bidirectional RNN focuses on the intrinsic details within these vertical segment regions from left-to-right and right-to-left.

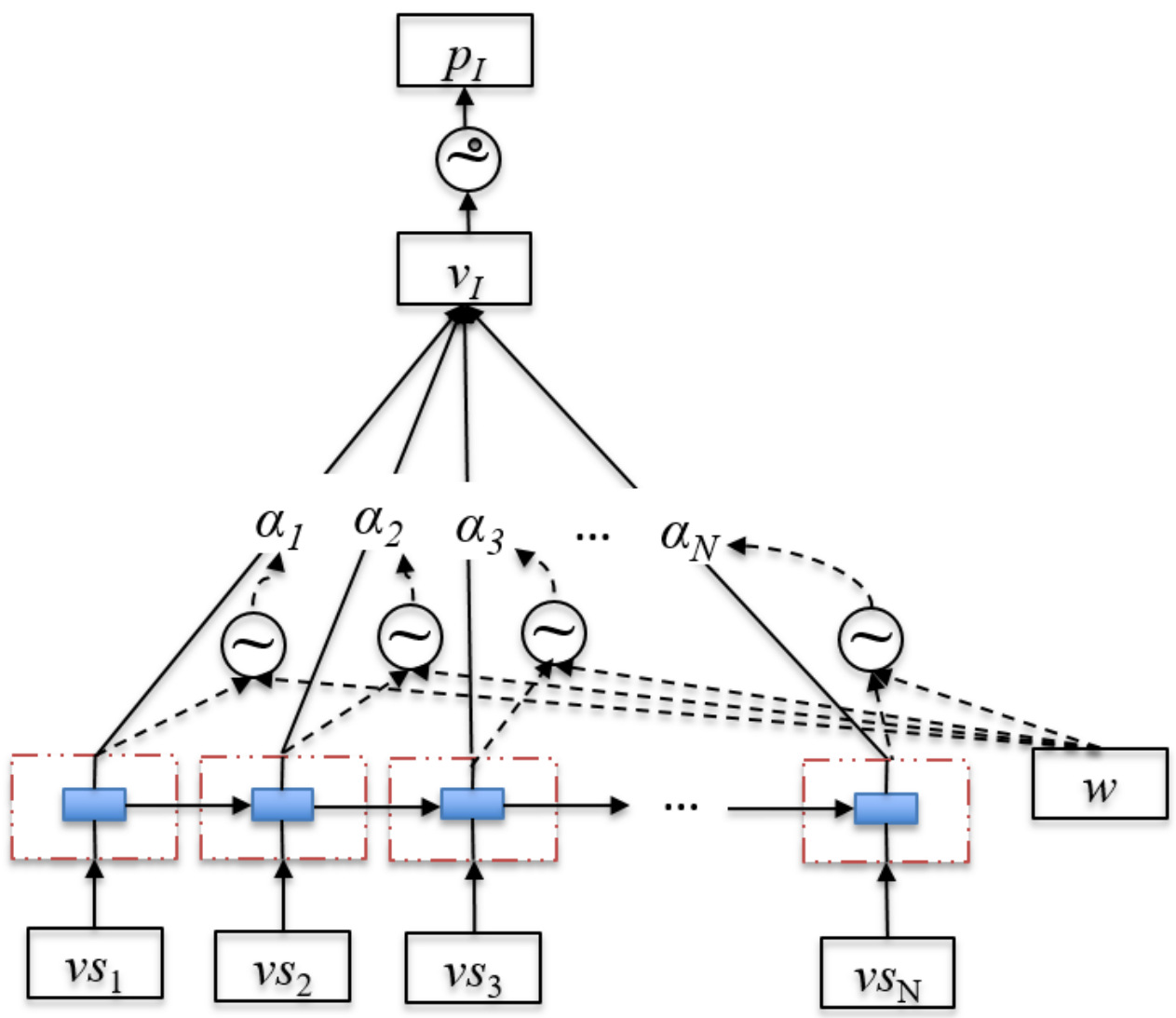

Fig. 5. Attention-based fusion network for epithelial image-level classification. The input sequences are fed to GRU cells. ~ denote a two layer neural network (NN) with hyperbolic tangent and Softmax activation functions, respectively to generate attentional weights. ~ denotes a single layer NN with Softmax activation function that produces the classification output.

The architecture of the proposed two-stage encoder model is given in Table I. The stage I encoder is built with first 87 layers of the DenseNet-121 model [24]. A max-pooling layer is added to this last layer such that the feature map has the height of 1. This can be considered as a feature sequence generated from left to right. Note that the convolutions always operate on local regions and hence are translationally invariant. Hence, the pixels in the feature maps from left-to-right corresponds to a local region in the original image (receptive field) from left-to-right. That is, the elements in the feature sequence are image descriptors in the same order. Importantly, they preserve the bottom-to-top spatial relationships in the original epithelium image. To further analyze this feature context, the generated feature sequence is fed to a stage II encoder built of RNNs. Specifically, we employed Bidirectional Long-Short-Term Memory (BLSTM) [25] networks to analyze and capture the long-term dependencies of the sequence from both directions. For the stage II encoder, two sets of BLSTM and single layer neural networks (NN) were appended to the last max-pooling layer of the stage I encoder. The final classification result is extracted from the logit vector of the last element in the output sequence generated at the stage II encoder. These logit vectors summarize the information of all the vertical segments and, when combined, form an information sequence that is fused to determine the image-level CIN classification.

Assuming an epithelial image with $N$ vertical segments $I_{vs_i}$, we have created logit sequence vectors $vs_i$ obtained with a segment-level sequence generator $f_s(\cdot\,;\theta)$:

$$vs_i = f_s(I_{vs_i};\theta) \tag{2}$$

where, $\theta$ represents the model parameters.

### C. Image-level Classification

The image-level classifier network is designed as an attention-mechanism based fusion network as shown in Fig. 5. The input sequences are picked up by a gated recurrent unit (GRU) [17], which tracks the state of the sequences with a gating mechanism. The output is a sequence vector that represents the image under test. We use a small classifier with an attentional weight for each GRU cell output to encode the sequence of the vertical segments as:

$$h_i = GRU(vs_i; h_{i-1}) \tag{3}$$

where $i \in [1, N]$ and $h_i$ is the hidden state that summarizes the information of the vertical segment $I_{vs_i}$.

The vertical segments may not contribute equally to epithelial image classification. We use an attention mechanism with a randomly initialized segment-level context vector $w$. This vector is used to generate the attentional weights $\alpha_i$ which analyze the contextual information and give a measure of importance of the vertical segments. The following equations explain the employed attention mechanism:

$$e_i = w^T \tanh(W_{vs}h_i + b_{vs}) \tag{4}$$

$$\alpha_i = \frac{exp(e_i)}{\sum_{i=1}^{N} exp(e_i)} \tag{5}$$

$$v_I = \sum_{i=1}^{N} \alpha_i h_i, \tag{6}$$

where $W_{vs}$ and $b_{vs}$ are trainable weights and bias. $v_I$ is the image feature vector that summarizes all the information of vertical segments in an epithelial image. The image-level classification is determined by:

$$p_I = softmax(W_0 v_I + b_0). \tag{7}$$

### D. Training

We trained the proposed networks independently with stratified K-fold cross validation split at the image-level. First the segment-level sequence generator is trained to generate the logit vectors of all the segments and then concatenated to form a sequence to further train the image-level classifier.

During segment-level sequence generation, the problem of class imbalance is solved by up-sampling the vertical segment images with image augmentations: randomly flipping vertically and horizontally, rotating with a range of 180 to -180 degree angles, changing hue, saturation, value and contrast, and applying blur and noise. The objective is to minimize the cross-entropy loss (equation 8) calculated directly from the vertical segment image and its restricted ground-truth label given by

$$L_k = -\sum_{vs} log\left(\frac{\exp(y_k)}{\sum_j \exp(y_j)}\right) \quad (8)$$

where $k$ is the class label of vertical segment image $vs$ and $y_k$ is the k^th^ label element value in the logit vector. We use ADADELTA [26] for optimization since it automatically adapts the learning rates based on the gradient updates. The initial learning rate set to 0.01.

TABLE II
CLASS LABEL DISTRIBUTION

| Class | Epithelial Images | | Segments | |
|---|---|---|---|---|
| | Count | % | Count | % |
| Normal | 244 | 53.8 | 6,836 | 57.7 |
| CIN1 | 57 | 12.6 | 1,433 | 12.1 |
| CIN2 | 79 | 17.5 | 2,039 | 17.2 |
| CIN3 | 73 | 16.1 | 1,546 | 13.0 |
| Total | 453 | 100.0 | 11,854 | 100.0 |

For image-level classification, we use the weighted negative log likelihood of correct labels to compute the cost function and back propagate the error to update the weights with a stochastic gradient descent (SGD) optimizer (learning rate was fixed at 0.0001). Training loss is given by:

$$L'_k = -q_k \sum_I log\,(p_{I_k}) \quad (9)$$

where $k$ is the class label of epithelial image $I$ and $q_k$ is the weight of the label $k$.

## III. EXPERIMENTS

We conducted experiments on our cervical histopathology image database to evaluate the effectiveness of the proposed classification model and compared its performance with other state-of-the-art methods.

### A. Dataset and Evaluation Metrics

For all the experiments, we use a dataset that contains 453 high-resolution cervical epithelial images extracted from 146 hematoxylin and eosin stained cervical histology WSIs. These WSIs were provided by Department of Pathology at the University of Oklahoma Medical Center in collaboration with the National Library of Medicine. They were scanned at 20X using Aperio ScanScope slide scanner and saved with the file extension *svs*. All 453 images have corresponding ground-truth labels. These annotations were carried out by an expert pathologist. The epithelial images have varying sizes which range from about 550 × 680 pixels (smallest) to 7500 × 1500 pixels (largest). This varying size affects the number of vertical segments generated from an image, typically ranging from 6 to 118. Though the vertical segments are generated such that the widths are 64 pixels wide, the height of these segments range from 160 to 1400 pixels. We address this problem by resizing the images to their median height: 704 pixels. This height was chosen empirically as a multiple of 32 in order to apply convolutions for feature extraction.

The segments were pre-processed such that they are RGB images of standard size: 64 × 704 × 3. We have created a total of 11,854 vertical segment images from 453 epithelial images. The class distribution of these data is shown in Table II.

There are two main challenges with this epithelial image dataset. First, the cervical tissues have irregular epithelium regions, with color variations, intensity variations, red stain blobs, variations in nuclei shapes and sizes, and noise and blurring effects created during image acquisition. These effects tend to have large inter- and intra-class variability across the four classes we seek to label. Second, even though our database is labelled by experts and may be considered of high quality, it is relatively small. This is a common and recognized problem in the biomedical image processing domain. Compared to large databases such as ImageNet [27], which has more than 14 million images with nearly 21 thousand classes, our data is very limited.

The scoring metrics used for the performance evaluation are precision (P), recall (R), F1-score (F1), classification accuracy (ACC), area under Receiver Operating Characteristic curve (AUC), average precision (AP) and Matthews correlation coefficient (MCC). Cohen's kappa score (κ) is used for the evaluation of the scoring schemes described in Section III.D. The percentage weighted average scores were reported due to the inevitable imbalance in the data distribution.

### B. Implementation Details

Although the entire DeepCIN model can be implemented end-to-end, we have split the process into two independent training steps. This model was chosen to overcome the GPU memory limitation to process these large input images and network architectures.

Details about the segment-level sequence generator network and image-level classifier network are given in Table I and Fig. 4, respectively. Both networks output four classes. The first network is trained with weak supervision to determine the logit sequence vectors of each vertical segment. The class outputs of the final network comprise our major concern.

A transfer learning technique was incorporated in the stage I encoder of the segment-level sequence generator. The convolution filters were initialized with ImageNet pre-trained weights and were left frozen since the encoder is built with initial layers of the DenseNet-121 model which presumably has weights already set to extract low-level image features such as edges, colors and curves. All the CNN layers are activated with the rectified linear unit (ReLU) function, and the single layer neural network followed by BLSTM layers in the stage II encoder, which does not impose any non-linearity. The latter network consists of GRU cells (with 128 hidden units), a two-layer neural network (NN) with hyperbolic tangent and Softmax activation functions, respectively, to generate attentional weights, and a single-layer NN with Softmax activation function to produce the classification output from the image feature vector.

TABLE III
ABLATION STUDY ON SEGMENT WIDTHS

| Segment width | P | R | F1 | ACC | AUC | AP | MCC |
|---|---|---|---|---|---|---|---|
| 32 | 82.9 | 82.3 | 81.2 | 82.3 | 93.5 | 85.3 | 72.3 |
| 64* | 88.6 | 88.5 | 88.0 | 88.5 | 96.5 | 91.5 | 82.0 |
| 128 | 85.3 | 85.6 | 84.9 | 85.6 | 95.9 | 89.8 | 77.1 |

TABLE IV
ABLATION STUDY ON STAGE I ENCODER MODELS

| Stage I Encoder | P | R | F1 | ACC | AUC | AP | MCC |
|---|---|---|---|---|---|---|---|
| DesnseNet-121* | 88.6 | 88.5 | 88.0 | 88.5 | 96.5 | 91.5 | 82.0 |
| ResNet-101 | 87.1 | 86.9 | 86.4 | 86.9 | 95.0 | 88.9 | 79.6 |
| Inception-v3 | 85.5 | 85.4 | 85.1 | 85.4 | 94.8 | 87.8 | 77.1 |

TABLE V
ABLATION STUDY ON STAGE II ENCODER MODELS

| Stage II Encoder | P | R | F1 | ACC | AUC | AP | MCC |
|---|---|---|---|---|---|---|---|
| BLSTM* | 88.6 | 88.5 | 88.0 | 88.5 | 96.5 | 91.5 | 82.0 |
| BLSTM + Attention | 87.9 | 87.6 | 87.7 | 87.6 | 95.2 | 88.9 | 80.1 |
| FC | 85.3 | 85.0 | 84.2 | 85.0 | 94.7 | 87.4 | 76.3 |

TABLE VI
ABLATION STUDY ON FUSION TECHNIQUES

| Fusion | P | R | F1 | ACC | AUC | AP | MCC |
|---|---|---|---|---|---|---|---|
| GRU | 86.3 | 86.1 | 85.6 | 86.1 | 96.3 | 90.4 | 78.0 |
| GRU+Attention* | 88.6 | 88.5 | 88.0 | 88.5 | 96.5 | 91.5 | 82.0 |
| Max vote | 87.6 | 87.2 | 87.0 | 87.2 | - | - | 79.9 |
| Avg vote | 88.0 | 87.6 | 87.4 | 87.6 | - | - | 80.6 |

We trained and validated the models using stratified 5-fold cross validation. We split training and validation data at the image level and maintained the same distribution across both the models. To address the class imbalance problem, we have up-scaled the less populated class images with image augmentations for the segment level sequence generation and in the image level classification we employed a weighted loss function.

Each individual fold for both the models were trained for 200 epochs with batch size of 56 with early stopping to avoid overfitting.

We implemented our localized vertical segment generation in MATLAB [28] running on an Intel Xeon CPU @ 2.10GHz which took 3.42 seconds on average to process one epithelial image. The deep learning models are trained under CUDA 10.2 and CuDNN v7.6 backend on an NVIDIA Quadro P4000 8GB GPU and 64GB RAM with a PyTorch v1.4 [29] framework. The time taken for validation is about 0.68 seconds per epithelial image. Thus, the entire DeepCIN pipeline takes 4.10 seconds on average to process and validate one epithelial image.

## C. Ablation Studies

In this section, we perform classifier ablation studies on the DeepCIN pipeline to understand its key aspects. The experiments include comparison with different segment widths, stage I encoder and stage II encoder variants, different fusion techniques and benchmark models.

The proposed model takes standard size image inputs. Resizing images will cause image distortions. We observe that this has a minor effect on the performance, expected since both the training and testing images are similarly resized which would result in the model's capability of handling such distortions. But the segment width is to some extent a free variable whose setting may modulate the amount of local spatial information contained in a vertical segment. Recognizing this, we experimented with segment widths of 32, 64 and 128. According to Table III, we observe that a segment width of 64 pixels is an optimal choice (in our experimental search space) compared to the segments with 32 pixels wide and 128 pixels wide.

The stage I encoder in the segment level sequence generator acts like a spatial feature extractor. Because our biomedical digital image environment is not data-rich for training deep learning models, we have experimented with various published models which have been pre-trained with the benchmark ImageNet database. Only a set of initial layers that extract low level features from the input image are considered in building the stage I encoder. We observed (from Table IV) that DenseNet-121 was better at extracting the crucial epithelial information, compared to ResNet-101 [30] and Inception_v3 [31] models. The DenseNet-121 model is better at feature reuse and feature propagation throughout the network with reduced parameters. Both DenseNet-121 and ResNet-101 are good at alleviating vanishing gradient problems, however DenseNet-121 with its feed-forward interconnections among layers helps in better feature understanding. Inception-v3 uses models that are wider rather than deeper to prevent overfitting with factorizing convolutions to reduce the number of parameters without compromising network efficiency.

The stage II encoder further encodes the feature sequence that is mapped from the translationally invariant feature information available from the stage I encoder. Our efforts to use bidirectional LSTM as a stage II encoder delivered better performance on the segment-level sequence generation that reflects on generating essential and better logit feature vectors. Table V shows that bidirectional analysis is enables understanding of the context of the feature information; this aided in up-sampling the segment data by flipping the input images horizontally. The use of attention was not helpful for understanding the feature sequence in the vertical segments with almost 1% decrease in performance across all the metrics (Table V). This indicates that the entire feature sequence is equally important to interpret the localized information, as shown by the equal distribution of attentional weights. The use of vanilla neural networks (fully connected layers) was comparatively less efficient because LSTMs contain internal state cells that act as long-term and short-term memory units and manage to learn by remembering the important information and forgetting the unwanted. Neural networks lack this ability and focus only on the very last input.

We observed that attentional weights help analyze the valuable information from the contribution of each segment towards the image-level classification. Table VI confirms this observation, showing nearly a 2% improvement in performance with inclusion of attention. Techniques like maximum voting and average voting of segment-level sequence generation results are simple and straight-forward, but fail provide the additional information about the localized segment data.

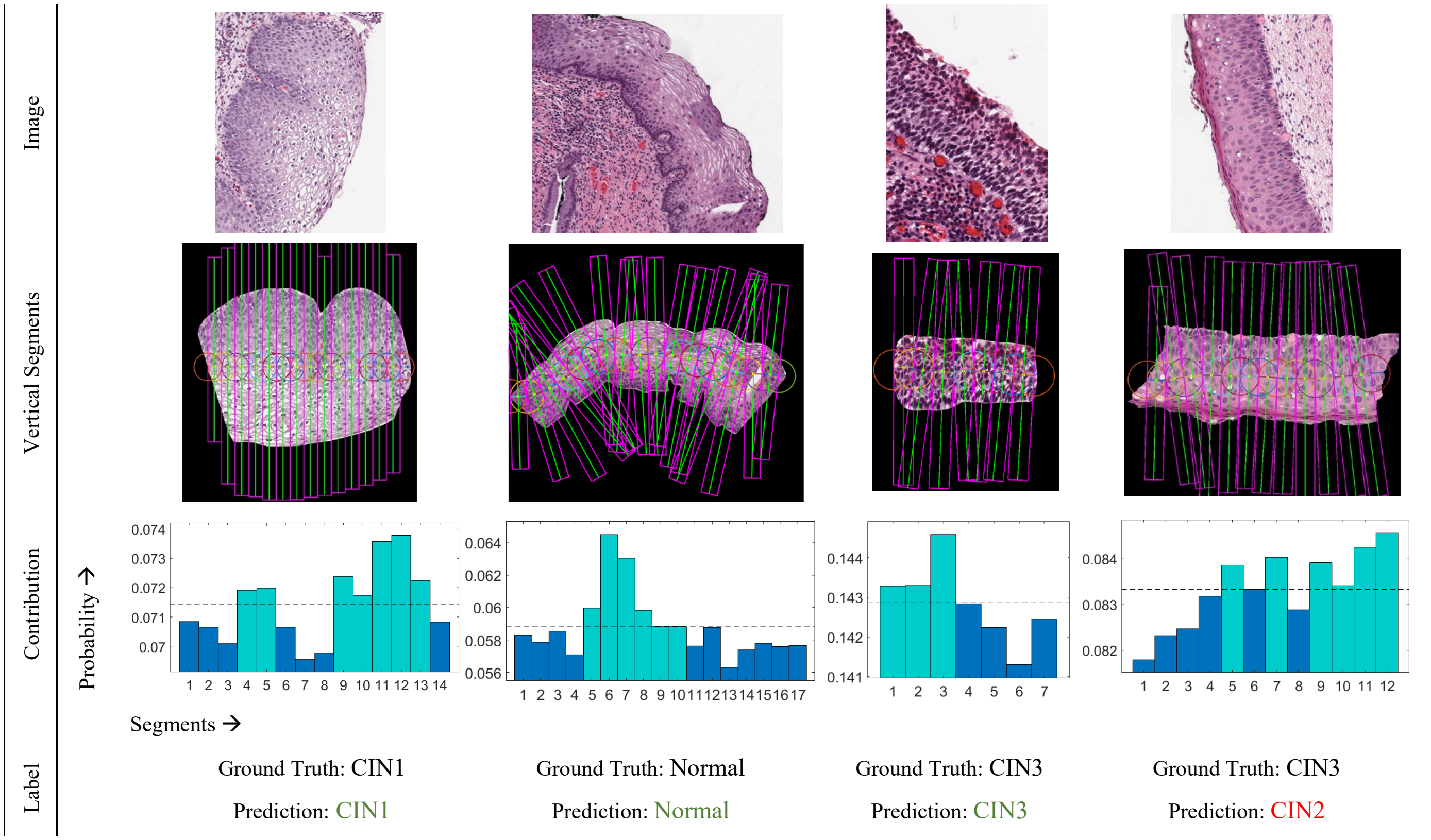

Fig. 6. Results of DeepCIN. From top to bottom, each column presents original image, localized vertical regions, contribution of segments within an image towards the image-level CIN classification (represented as probability distribution over the segments (attentional weights), the dotted lines indicate mean value and segments above the mean value, highlighted in green, are contributing the most), and corresponding ground truth and prediction labels, respectively.

TABLE VII
COMPARISON WITH STATE OF THE ART MODELS

| Model | P | R | F1 | ACC | AUC | AP | MCC |
|---|---|---|---|---|---|---|---|
| Guo *et al.* [6] | 67.5 | 73.3 | 69.4 | 73.4 | - | - | 56.5 |
| AlMubarak *et al.* [11] | 66.1 | 75.6 | 70.4 | 75.5 | 90.9 | 78.1 | 60.3 |
| Ours* | 88.6 | 88.5 | 88.0 | 88.5 | 96.5 | 91.5 | 82.0 |

## IV. PERFORMANCE OF DEEPCIN

We finally compare the performance of the proposed model with the state-of-the-art CIN classification models. The models used for the comparison are proposed by Guo *et al.* [6] and AlMubarak *et al.* [11]. The best model of Guo *et al.* [6], linear discriminant analysis, was trained with 27 handcrafted features extracted from vertical image segments. The epithelium was split into 10 equal parts to create these segments and fusion was performed through a voting scheme. AlMubarak *et al.* [11] used the same vertical segments and divided them into three sections: top, middle and bottom. 64 × 64 size Lab color space image patches were extracted to train three CNN models. The resulting confidence values from these sections were treated as features, and the 27 features were concatenated to form a hybrid approach for training an SVM classifier. The final classifiers of both these models were trained with a leave-one-out approach.

For a direct comparison, we have retrained [6] and [11] models on the 453 high-resolution epithelial histopathology image data. Table VII shows that the proposed model performs best for the CIN classification task. Additionally, our model provides the significance of individual local regions towards the whole image classification. Results for sample images from the proposed DeepCIN model are shown in Fig. 6. The distribution of the entire data and the predictions for all 5-folds is depicted in the Sankey diagram in Fig. 7, which shows the proportion of images that are correctly classified and misclassified. Image samples belonging to the CIN1 class were mostly misclassified as normal class. Two reasons may explain this: 1) CIN1 images closely resemble normal images; 2) the number of CIN1 class images is small, relative to the number of Normal class images.

TABLE VIII
CIN CLASSIFICATION RESULTS WITH DIFFERENT SCORING SCHEMES

| Scoring Scheme | P | R | F1 | ACC | AUC | AP | MCC | κ |
|---|---|---|---|---|---|---|---|---|
| Exact class label | 88.6 | 88.5 | 88.0 | 88.5 | 96.5 | 91.5 | 82.0 | 81.5 |
| CIN vs Normal | 94.6 | 94.1 | 94.0 | 94.1 | 93.8 | 97.7 | 88.5 | 87.9 |
| CIN3-CIN2 vs CIN1-Normal | 96.8 | 96.7 | 96.7 | 96.7 | 96.0 | 98.9 | 92.7 | 92.5 |
| CIN3 vs CIN2-CIN1-Normal | 96.2 | 96.0 | 96.0 | 96.0 | 88.4 | 98.3 | 85.3 | 84.8 |
| Off-by-one | - | - | - | 98.9 | - | - | - | - |

As an extension, we have tabulated the performance model with exact class labels, CIN versus Normal, CIN3-CIN2 versus CIN1-Normal, CIN3 versus CIN2-CIN1-Normal, and off-by-one class (Table VIII). For the exact class label scheme, the predicted class label should exactly match the expert ground-truth class label. The CIN versus Normal scheme is an abnormal-normal grouping of the predicted labels. The CIN3-CIN2 versus CIN1-Normal and CIN3 versus CIN2-CIN1-Normal interclass grouping schemes resemble the clinical decisions for treatment. The Off-by-one scheme emphasizes the possible disagreement between the expect pathologists while labelling the CIN class which is usually observed to be one grade off [32].

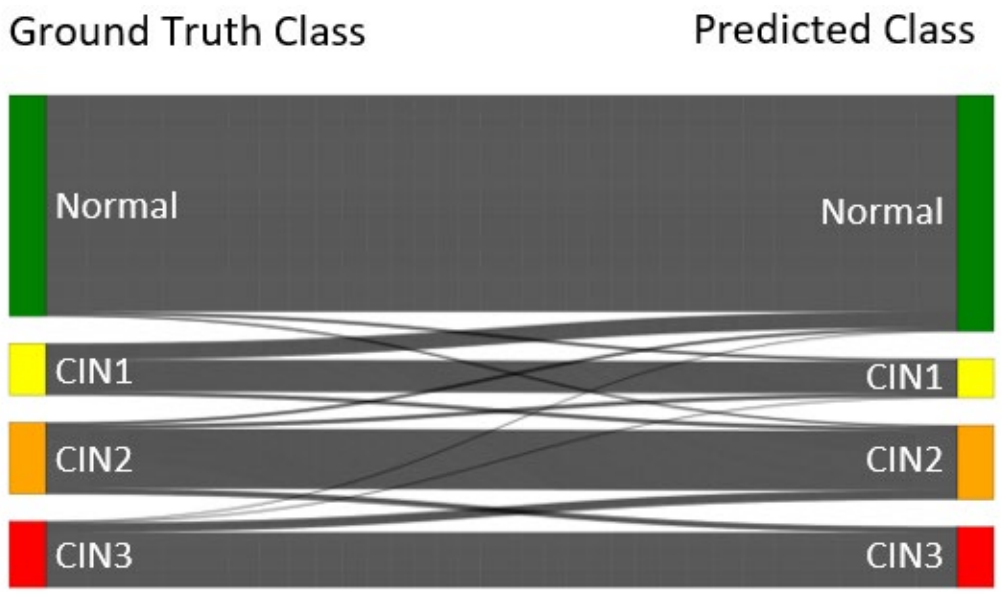

Fig. 7. Sankey diagram – based on the combined test results from the 5-fold cross-validation. The height of each bar is proportional to the number of samples corresponding to each class.

## V. Discussion

The main objective of the DeepCIN model is to classify the high-resolution epithelium images into normal or precancerous transformation of cells of the uterine cervix. We generate classification results by fusing localized information, forming a sequence of logit feature vectors in the same order of the vertical segments from the epithelium image. The number of vertical segments created varies since the epithelium images have arbitrary shapes. Traditional neural networks are limited to fixed-length input, but RNNs have the capability to read varying input sequences along with memorization. We employ a GRU to read the arbitrarily shaped input sequences. GRU with attention helps in better understanding the differentially informative localized data. Unlike the stage II encoder from the segment-level sequence generator, incorporation of attention helped the model to better fuse the segment data and identify localized regions that are significantly important in the classifying the epithelial image.

It is now four decades since Marsden Scott Blois presented a paradigm for medical information science to distinguish domains in medicine in which humans are essential from those in which computation is essential and computers are likely to play a primary role [33]. He emphasized the importance of human judgment in the former domain, which includes most of clinical medicine, but does not include the evaluation and interpretation of physiological parameters, for example blood gases, which is the proper domain of computers. With regard to the Blois paradigm, we propose that computer processing of histopathology images falls within the computational domain, and computers are likely to play a primary role.

## VI. Conclusion

In this study, we address the CIN classification problem by focusing on localized epithelium regions. The varying atypical nuclei density which is crucial in CIN determination is better analyzed by sequence mapping of the deep learning features. This sequence is interpreted in both directions under weak supervision with the long-term and short-term memory of the feature information. We employed an attention-based fusion approach to carry out an image-level classification. This hierarchical approach not only produces the image-level CIN classification labels but also provides the contribution of each individual vertical segment of the epithelium towards the whole image classification. We conjecture that this information highlights the highest-risk areas; this serves as an automated check for the pathologist's assessment.

We observed that our proposed model, DeepCIN, has out-performed state-of-the-art models in classification accuracy. The final image-level classification accuracies and Cohen's kappa score are {88.5% (± 2.2%), 81.5%}, {94.1% (± 2.0%), 87.9%}, {96.7% (±1.6%), 92.5%}, {96.0% (±1.7%), 84.8%}, and {98.9% (± 0.0%), -} for exact class label, CIN versus Normal, CIN3-CIN2 versus CIN1-Normal, CIN3 versus CIN2-CIN1-Normal and leave-one-out schemes, respectively. These results significantly exceed the variability of community pathologists when measured against the gold standard, and are in the range of inter-pathologist variability for expert pathologists as measured by the κ statistics.

Limitations of this work include use of a database that is not publicly available, which precludes validation by other researchers. Ground truth for the entire set was based on only one expert pathologist. Part of the set was scored by two pathologists; accuracies obtained for the two sets are similar. Future work could improve results by including more annotated image data with balanced class distribution for training. There is also a possibility for improvements if the entire model could be trained end-to-end, which requires greater GPU resources. Our future research will focus on WSI-level classification with end-to-end automation which combines the proposed model with our previous work on automated epithelium segmentation [34], and automated nuclei detection [35] for extracting enhanced feature information.